\documentclass[aps,prd,
amsmath,floats,floatfix, twocolumn, superscriptaddress,nofootinbib,showpacs]{revtex4-1}
\usepackage[usenames, dvipsnames]{color}
\usepackage[colorlinks, pdfborder={0 0 0}, plainpages=false,urlcolor=NavyBlue,citecolor=Green]{hyperref}
\usepackage{graphicx}
\usepackage{xspace}
\usepackage[usenames,dvipsnames]{color}
\usepackage{amssymb}
\usepackage[normalem]{ulem} 
\usepackage{letltxmacro}
\usepackage{amsfonts}
\usepackage{amsmath}
\usepackage{amssymb}
\usepackage{bm}
\usepackage{dcolumn}
\usepackage{epsfig}
\usepackage{graphicx}
\usepackage{graphics}
\usepackage[latin1]{inputenc}
\usepackage{latexsym}
\usepackage{rotating}
\usepackage{hyperref}
\usepackage[usenames]{color}
\usepackage{float}
\usepackage{xspace} 
\usepackage{mathrsfs}
\usepackage{subfigure}
\usepackage{multirow}
\usepackage{url}
\usepackage{times}

\def\prd{{\it Phys. Rev.} D~}

\def\apjl{{\it Astrophys. J.} Lett~}
\def\apj{{\it Astrophys. J.}}
\def\Msun{M_\odot}

\def\mnras{{\it MNRAS~}}

\def\lesssim{\mathrel{\hbox{\rlap{\hbox{\lower4pt\hbox{$\sim$}}}\hbox{$<$}}}}
\def\gtrsim{\mathrel{\hbox{\rlap{\hbox{\lower4pt\hbox{$\sim$}}}\hbox{$>$}}}}
\def\alt{\mathrel{\hbox{\rlap{\hbox{\lower4pt\hbox{$\sim$}}}\hbox{$<$}}}}
\def\agt{\mathrel{\hbox{\rlap{\hbox{\lower4pt\hbox{$\sim$}}}\hbox{$>$}}}}
\def\apjs{{\it ApJS}}

\def\gta{\ifmmode {\mathbin{\lower 3pt\hbox   
    {$\,\rlap{\raise 5pt\hbox{$\char'076$}}\mathchar"7218\,$}}}
    \else {${\mathbin{\lower 3pt\hbox
    {$\rlap{\raise 5pt\hbox{$\char'076$}}\mathchar"7218\,$}}}
    $}\fi}
\def\lta{\ifmmode {\,\mathbin{\lower 3pt\hbox   
    {$\,\rlap{\raise 5pt\hbox{$\char'074$}}\mathchar"7218\,$}}}
    \else {${\mathbin{\lower 3pt\hbox
    {$\rlap{\raise 5pt\hbox{$\char'074$}}\mathchar"7218\,$}}}
    $}\fi}

\newcommand{\beq}{\begin{equation}}
\newcommand{\eeq}{\end{equation}}
\newcommand{\bea}{\begin{eqnarray}}
\newcommand{\eea}{\end{eqnarray}}

\newcommand{\NCSA}{\affiliation{NCSA, University of Illinois at Urbana-Champaign, Urbana, Illinois, 61801, USA}}
\newcommand{\ANCSA}{\affiliation{Department of Astronomy, University of Illinois at Urbana-Champaign, Urbana, Illinois, 61801, USA}}

\begin{document}

\title{Deep Neural Networks to Enable Real-time Multimessenger Astrophysics}


\author{Daniel George}\ANCSA \NCSA 
\author{E.~A. Huerta}\NCSA


\begin{abstract}
    Gravitational wave astronomy has set in motion a scientific revolution. To further enhance the science reach of this emergent field, there is a pressing need to increase the depth and speed of the gravitational wave algorithms that have enabled these groundbreaking discoveries. To contribute to this effort, we introduce \texttt{Deep Filtering}, a new highly scalable method for end-to-end time-series signal processing, based on a system of two deep convolutional neural networks, which we designed for \textit{classification} and \textit{regression} to rapidly detect and estimate parameters of signals in highly noisy time-series data streams. We demonstrate a novel training scheme with gradually increasing noise levels, and a transfer learning procedure between the two networks. We showcase the application of this method for the \textit{detection} and \textit{parameter estimation} of gravitational waves from binary black hole mergers. Our results indicate that \texttt{Deep Filtering} significantly outperforms conventional machine learning techniques, achieves similar performance compared to matched-filtering while being several orders of magnitude faster thus allowing real-time processing of raw big data with minimal resources. More importantly, \texttt{Deep Filtering} extends the range of gravitational wave signals that can be detected with ground-based gravitational wave detectors. This framework leverages recent advances in artificial intelligence algorithms and emerging hardware architectures, such as deep-learning-optimized GPUs, to facilitate real-time searches of gravitational wave sources and their electromagnetic and astro-particle counterparts.

\end{abstract}


\maketitle

\section{Introduction}
\label{intro}

Gravitational wave (GW) astrophysics is by now a well established field of research. The advanced Laser Interferometer Gravitational wave Observatory (aLIGO) detectors have detected four significant GW events, consistent with Einstein's general relativity predictions of binary black hole (BBH) mergers~\cite{DI:2016,secondBBH:2016,bbhswithligo:2016,thirddetection}. These major scientific breakthroughs, worthy of the 2017 Nobel Prize in Physics, have initiated a new era in astronomy and astrophysics. By the end of aLIGO's second discovery campaign, referred to as O2, the European advanced Virgo detector~\cite{Virgo:2015} joined aLIGO, establishing the first, three-detector search for GW sources in the advanced detector era. We expect that ongoing improvements in the sensitivity of aLIGO and Virgo within the next few months will continue to increase the number and types of GW sources.

GW astrophysics is a multidisciplinary enterprise. Experimental and theoretical physics, cosmology, fundamental physics, high performance computing (HPC) and high throughout computing have been combined into a coherent program to revolutionize our understanding of the Universe. For instance, at the interface of HPC and theoretical physics, numerical relativity (NR) simulations of Einstein's field equations are extensively used to validate the astrophysical nature of GW transients~\cite{NRI:2016}. Furthermore, NR simulations of binary neutron star (BNS) mergers, neutron star-black hole (NSBH) mergers, core collapse supernovae and other massive, relativistic systems provide key physical insights into the physics of systems that are expected to generate electromagnetic (EM) and astro-particle counterparts~\cite{sum:2009CQGra,2014CQGra..31a5005M,2016PhRvD..93l4062H,2014PhRvD..90d4001A,2017JCoPh.335...84K,2013ApJ...767..124N}.

Ongoing discovery campaigns with GW detectors and astronomical facilities~\cite{DES:2016MNRAS,fermi:2013ApJS,LSST:2002SPIE,euclid:2013LRR,wfirst:2015JPhCS,D12:2016} have already led to \textit{multimessenger} observations of GW events and their EM counterparts~\cite{BNSdet:2017,scenarioligo:2016LRR,Singer:2014ApJ}. These complementary observations have provided new and detailed information about the astrophysical origin, and cosmic evolution of ultra compact objects~\cite{mma:2017,Eichler:1989,Paczynski:1986,Narayan:1992,Kochanek:1993mw,sum:2009CQGra,phi:2009astro2010S}. The time sensitive nature of these analyses requires algorithms that can detect and characterize GW events in real-time~\cite{2016ApJ...820....7L}. 

aLIGO's flagship matched-filtering searches have been very successful at identifying and characterizing GW transients~\cite{2016CQGra..33u5004U,2012ApJ...748..136C,D6:2016,corn:2015CQGra}. Looking ahead in the near future, GW discovery campaigns will be longer, and data will be gathered by a network of interferometers in several continents. In anticipation for this scenario, LIGO scientists now exploit state-of-the-art HPC facilities to increase the pool of computational resources to carry out for large scale GW data analysis. To maximize the science we can extract from GW observations, it is essential to cover a deeper parameter space of astrophysically motivated sources, i.e., we need to increase the dimensionality of existing GW searches from 3-dimensions (3D) to 9D\footnote{9D: component masses, eccentricity, and two (3D) vectors describing the spin of each binary component.}. Furthermore, accelerating parameter estimation algorithms, which typically last from several hours to a few days, is no trivial task since they have to sample a 15D parameter space~\cite{Smith:2016PhRvD}. This is a grand computational challenge given the compute intensive nature of large scale GW searches~\cite{2016PhRvD..94b4012H}.  

To start addressing these pressing issues, we introduce \texttt{Deep Filtering}, a new machine (deep) learning algorithm, based on deep neural networks (DNNs)~\cite{DL-Nature} to directly process highly noisy time-series data for both classification and regression. \texttt{Deep Filtering} consists of two deep convolutional neural networks~\cite{ConvNet} that directly take time-series inputs and are capable of detecting and characterizing signals whose peak power is significantly weaker than that of the background noise. In this foundational article, we carry out a systematic assessment of DNNs trained to cover the stellar-mass, BBH parameter-space, where ground-based GW detectors are expected to have the highest detection rate~\cite{bel:2016Na}. As a first step, to construct and validate \texttt{Deep Filtering}, we have used a dataset of inspiral-merger-ringdown (IMR) BBH waveforms for training~\cite{Tara:2014}. 

As discussed in~\cite{2016PhRvD..94b4012H}, the computational cost of matched-filtering searches increases significantly when targeting GW sources that span a higher dimensional parameter space. In contrast, when using deep learning, all the intensive computation is diverted to the one-time training stage, after which the datasets can be discarded, i.e., the size of template banks that describe the GW signals we search for present no limitation when using deep learning. Indeed, it is preferable to use large datasets of GW signals for the one-time training stage to cover as deep a parameter space as possible. With existing computational resources on supercomputers such as Blue Waters, we estimate that it would possible to finish training the DNNs on templates across 10 or more dimensions of parameters within a few weeks.

The main objective in developing \texttt{Deep Filtering} is to enhance existing, low latency GW detection algorithms to enable deeper and faster GW searches. We envision using \texttt{Deep Filtering} to identify and rapidly constrain the astrophysical parameters of GW transients. This real-time analysis would then be followed up by existing LIGO pipelines focusing on a narrow region of GWs' higher dimensional parameter space. A targeted search of this nature will significantly reduce the size of multi-dimensional template banks, enabling the use of established matched-filtering searches at a fraction of their computational cost to quantify the significance of new GW detections. This approach would combine the best of two approaches: the scalable, multidimensional nature of neural networks with the sophistication of LIGO detection pipelines. To accomplish this, we are working with the developers of \texttt{PyCBC}~\cite{2016CQGra..33u5004U} to implement \texttt{Deep Filtering} as a module to increase the depth and speed of this pipeline.

The results we present in this article confirm that DNNs are ideal tools for future GW analysis. We have found that DNN are able to \textit{interpolate} between waveform templates,  in a similar manner to Gaussian Process Regression (GPR)~\footnote{GPR~\cite{2003itil.book.....M,gpr:2016PhRvD,moore:2014gpr} is a statistical tool that can serve as a probabilistic interpolation algorithm providing information about the training set of NR simulations needed to accurately describe a given parameter-space and generates interpolated waveforms that match NR counterparts above any given accuracy.}, and to \textit{generalize} to new classes of signals beyond the templates used for training. Furthermore, our DNNs can be evaluated faster than real-time with a single CPU, and very intensive searches over a broader range of signals can be easily carried out with one dedicated GPU. The intelligent nature of deep learning would allow automated learning of persistent and transient characteristics of noises inherent to the detectors, while incorporating real-time data quality information. This analysis, combined with recent work to understand and characterize aLIGO non-Gaussian noise transients~\cite{spy:2016arXiv,Geo:2017b}, strongly suggests that it is feasible to create a single efficient pipeline to perform all tasks---identifying the presence or absence of GW signals, classifying noise transients, and reconstructing the astrophysical properties of detected GW sources. Furthermore, since this technique can be applied to other types of raw time-series data, similar DNNs can be used to process telescope data, thus paving a natural path to realizing real-time multimessenger astrophysics with a unified framework.

As NR continues to shed light into the physics of GW sources\cite{NRI:2016}, we will rely on an extensive exploitation of HPC resources to obtain NR waveforms to train our DNN algorithm. At the same time, we are using HPC facilities to carry out large scale parameter sweeps to find optimal DNNs for GW detection and parameter estimation. The approach we discuss here employs recent advances in artificial intelligence algorithms, by computer scientists and industries, for accelerating scientific discovery by enhancing the use of traditional HPC resources, while allowing us to exploit emerging hardware architectures such as deep-learning-optimized Graphics Processing Units (GPUs)~\cite{cuDNN}, Application-Specific Integrated Circuits (ASICs)~\cite{TensorFlow}, Field-Programmable Gate Arrays (FPGAs)~\cite{FPGA}, quantum computers~\cite{quantum} and brain-like neuromorphic chips~\cite{TrueNorth-Science}. This approach may provide the needed platform to address common challenges on large scale data analytics on disparate fields of research to effectively consolidate different windows of observation into the Universe.

This article is organized as follows: Section~\ref{dnn} provides a comprehensive overview of artificial neural networks and deep learning, particularly focusing on convolutional neural networks in the context of time-series signal processing. In Section~\ref{meth}, we describe our assumptions, datasets, and procedure to construct the DNN-based GW analysis pipeline. We report the results of our analysis in Section~\ref{result}. In Section~\ref{disc}, we discuss its immediate applications, and their implications for GW astrophysics missions, along with scope for improvements. We summarize our findings and outline its broader impact in Section~\ref{conc}.


\section{Deep Neural Networks}
\label{dnn}

In this section we provide a brief overview of the main concepts of deep learning, including machine learning, artificial neural networks, and convolutional neural networks in the context of time-series signal processing.

The vast majority of algorithms are designed with a specific task in mind. They require extensive modifications before they can be re-used for any other task. The term machine learning refers to a special class of algorithms that can \textit{learn} from examples to solve new problems without being explicitly re-programmed. This enables cross-domain applications of the same algorithm by training it with different data~\cite{DL-Book}. More importantly, some of these algorithms are able to tackle problems which humans can solve intuitively but find difficult to explain using well-defined rules, hence they are often called ``artificial intelligence''~\cite{DL-Book}.

The two main categories of machine learning are supervised and unsupervised learning. In supervised learning, the algorithm learns from some data that is correctly labeled, while unsupervised learning algorithms have to make sense of unstructured and unlabeled data~\cite{DL-Review}. We will be focusing on an application of supervised learning in this work, where we use labeled data obtained from physics simulations to train an algorithm to detect signals embedded in noise and also estimate multiple parameters of the source.

Although traditional machine learning algorithms have been successful in several applications, they are limited in their ability to deal directly with raw data. Often the data has to be simplified manually into a representation suitable for each problem. Determining the right representation is extremely difficult and time-consuming, often requiring decades of effort even for domain experts, which severely limits the applicability of these algorithms~\cite{DL-Book}.

Representation learning is a subset of machine learning which aims to resolve this issue by creating algorithms that can learn by themselves to find useful representations of the raw data and extract relevant features from it automatically for each problem~\cite{RL-Review}. Here, we are focusing on a special type of representation learning called deep learning.

\subsection*{Deep Learning}

Deep learning is a new subfield of machine learning, which resolves this difficulty of feature engineering with algorithms that learn by themselves to find useful representations of the raw data, and extract multiple levels of relevant features from it automatically for each problem. This is achieved by combining a computational architecture containing long interconnected layers of ``artificial neurons'' with powerful learning (optimization) algorithms~\cite{DL-Nature,DL-Book}. These deep artificial neural networks (DNNs) are able to capture complex non-linear relationships in the data by composing hierarchical internal representations, all of which are learned automatically during the training stage. The deepest layers are able to learn highly abstract concepts, based on the simpler outputs of the previous layers, to solve problems that previously required human-level intelligence~\cite{DL-Review}.  

Various factors including the exponential growth of computational resources (especially GPUs), availability of massive amounts of data, and the development of new algorithmic techniques and software have recently contributed to make deep learning very successful in commercial applications, thus revolutionizing multiple industries today. The state-of-the-art algorithms for image processing, speech recognition, natural language understanding are all based on deep learning. DNNs power many of the technologies routinely used by us including search engines (Google, Bing), voice recognition, personal assistants (Siri, Cortana, Google assistant), text prediction on mobile keyboards, real-time face detection on cameras, face recognition (e.g. face-tagging in Facebook), language translation (Google Translate), text-to-speech synthesis~\cite{WaveNet}, recommendations on Amazon, and automatic captioning on YouTube, to name a few~\cite{BigDataAI}.

\subsection*{Artificial Neural Networks}

Artificial neural networks (ANN), the building blocks of DNNs, are biologically-inspired computational models that have the capability to learn from observational data~\cite{DNN-Book}. The fundamental units of neural networks are artificial neurons (loosely modeled after real neurons~\cite{ann2003}), which are based on perceptrons introduced by Rosenblatt in 1957~\cite{Perceptron}. A perceptron takes a vector of inputs ($\vec{x}$) and computes a weighted output with an offset known as bias. This can be modeled by the equation $f(\vec{x})=\vec{w} \cdot \vec{x}+b$, where the weights ($\vec{w}$) and bias ($b$) are learned through training.

Minsky and Papert showed in their 1969 book Perceptrons~\cite{MinskyBook} that a single perceptron has many limitations. Unfortunately, this led to a decline in the popularity of all neural networks in the following decades~\cite{DL-Review}. However, it was later found that these limitations can be overcome by using multiple layers of inter-connected perceptrons to create ANNs. The universality theorem~\cite{UnivTheorem} proves that ANNs with just three layers (one hidden layer) can model any function up to any desired level of accuracy.

Multilayer perceptrons are also known as feed-forward neural networks because information is propagated forward from the input layer to the output layer without internal cycles (i.e no feedback loops)~\cite{DL-Book}. While potentially more powerful cyclic architectures can be constructed, such as Recurrent Neural Networks (RNNs), we will be focusing mainly on simple feed-forward neural networks in this article. 

\begin{figure}
	\centering
	\includegraphics[width=.28\textwidth]{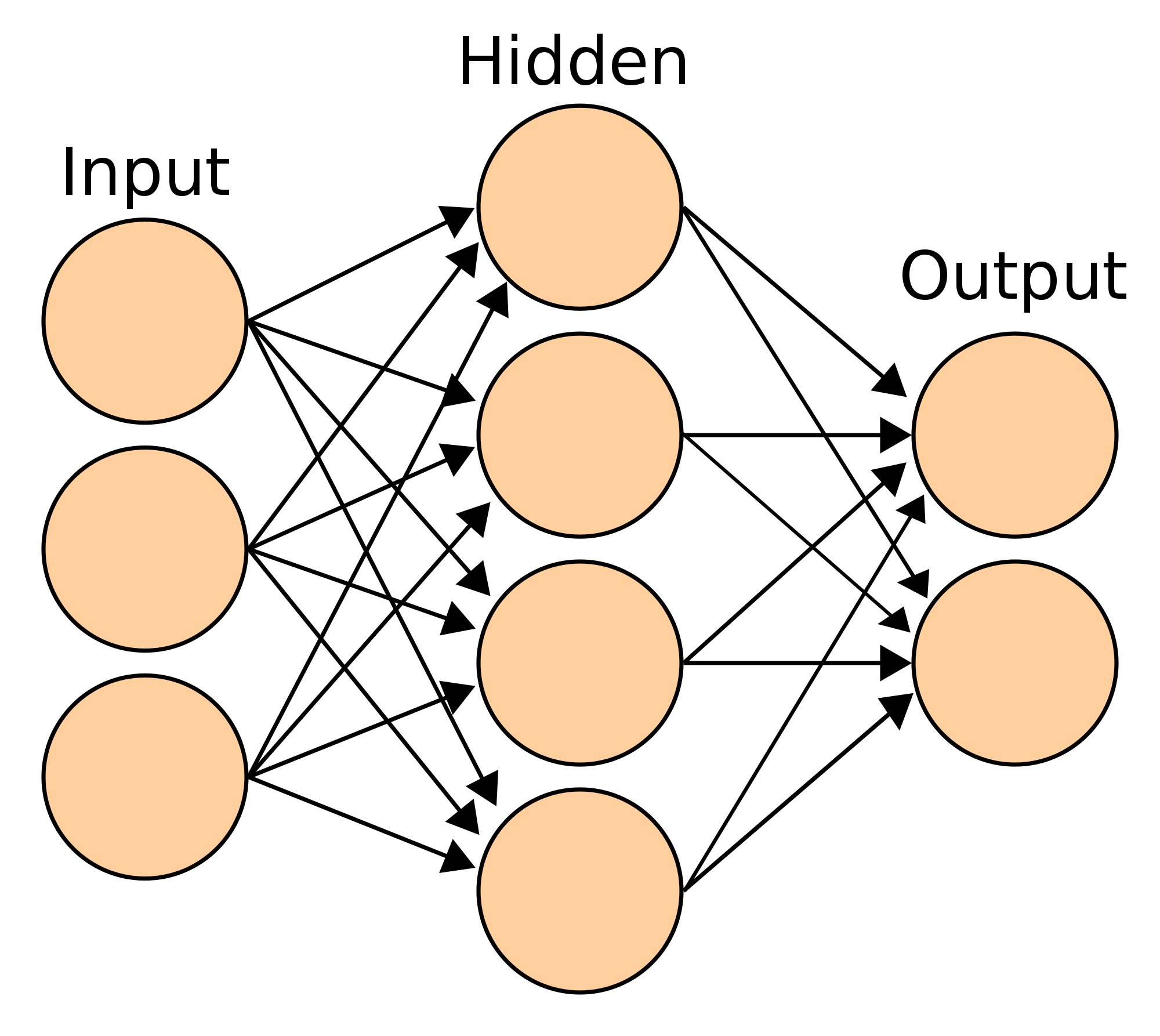}
	\caption{An Artificial Neural Network (ANN) or multilayer perceptron with one hidden layer is depicted~\cite{picANN}. The circles represent neurons and arrows represent connections (weights) between neurons. Note that each neuron has only a \textit{single} output, which branches out to connect with neurons in the next layer.
	}
	\label{fig:NN}
\end{figure}

An ANN usually has an input layer, one or more hidden layers, and an output layer (shown in Figure~\ref{fig:NN}). A non-linear ``activation'' function is applied to the output of each of the hidden layers. Without this non-linearity, using multiple layers would become redundant, as the network will only be able to express linear combinations of the input. The most commonly used non-linear activation functions are the logistic sigmoid, hyperbolic tan, and the rectified linear unit (also called ReLU or ramp). It has been empirically observed that the ramp produces the best results for most applications~\cite{ReLU} . This function is mathematically expressed as $max(0,x)$.

The key ingredient that makes ANNs useful is the learning algorithm. Almost all neural networks used today are trained with variants of the back-propagation algorithm based on the steepest descent method~\cite{DL-Review}. The idea is to propagate errors backward from the output layer to the input layer after each evaluation of a neural network, in order to adjust the weights of each neuron so that the overall error is reduced in a supervised learning problem~\cite{BackProp}. The weights of an ANN are usually initialized randomly to small values and then back-propagation is performed over multiple rounds, known as epochs, until the errors are minimized. Stochastic gradient descent with mini-batches~\cite{SGD} has been the traditional method used for back-propagation. This technique uses an estimate of the gradient of the error over subsets of the training data in each iteration to change the weights of the ANN. The magnitude of these changes is determined by the ``learning rate''. New methods with variable learning rates such as ADAM (Adaptive Momentum Estimation) are becoming more popular and have been shown empirically to achieve better results more quickly~\cite{ADAM}.

\subsection*{Convolutional Neural Networks}

A convolutional neural network (CNN), whose structure is inspired by studies of the visual cortex in mammals~\cite{DL-Book}, is a type of feed-forward neural network. First developed by Fukushima for his Neocognitron~\cite{Neocognitron}, they were successfully combined with back-propagation by LeCun~\cite{ConvNet} in the 1980s, for developing a highly accurate algorithm for recognizing handwritten digits.  The exceptional performance of Alex Krizhevsky's entry based on CNNs, which won the ImageNet competition by a huge margin in 2012~\cite{AlexNet}, has sparked the current interest in these networks especially in the field of computer vision. CNNs have been most effective for image and video processing. They have been shown to approach or even surpass human-level accuracy at a variety of constrained tasks such as hand-writing recognition, identifying objects in photos, tracking movements in videos etc.~\cite{DL-Nature}.

The introduction of a ``convolution layer'', containing a set of neurons that share their weights, is the critical component of these networks. Multiple convolution layers are commonly found in DNNs, with each having a separate set of shared weights that are learned during training. The name comes from the fact that an output equivalent to a convolution, or sometimes cross-correlation~\cite{DL-Book}, operation is computed with a kernel of fixed size. A convolutional layer can also be viewed as a layer of identical neurons that each ``look'' at small overlapping sections of the input, defined as the receptive field. 

The main advantage of using these layers is the ability to reduce computational costs by having shared weights and small kernels, thus allowing deeper networks and faster training and evaluation speeds. Because of the replicated structure, CNNs are also able to automatically deal with spatially translated as well as (with a few modifications~\cite{DL-Nature}) rotated and scaled signals. In practice, multiple modules each consisting of a sequence of convolution and pooling (sub-sampling) layers, followed by a non-linearity, are used. The pooling layers further reduces computational costs by constraining the size of the DNN, while also making the networks more resilient to noise and translations, thus enhancing their ability to handle new inputs~\cite{DL-Nature}. Dilated convolutions~\cite{dilatedCNN} is a recent development which enables rapid aggregation of information over larger regions by having gaps within each of the receptive fields. In this study, we focus on CNNs as they are the most efficient DNNs on modern hardware, allowing fast training and evaluation (inference).

\subsection*{Time-series Analysis with Convolutional Neural Networks}

Conventional methods of digital signal processing such as matched-filtering (cross-correlation or convolution against a set of templates)~\cite{owen:1999PhRvD..60b2002O} in time-domain or frequency-space are limited in their ability to scale to a large parameter-space of signal templates, while being too computationally intensive for real-time parameter estimation analysis~\cite{Smith:2016PhRvD}. Signal processing using machine learning in the context of GW astrophysics is an emerging field of research~\cite{bambiann:2015PhRvD,bambi:2012MNRAS,DBNN,jade1:2016,jade:2015CQGra,spy:2016arXiv,DeepTransferLearning}. These traditional machine learning techniques, including shallow ANNs, require ``handcrafted'' features extracted from the data as inputs rather than the raw noisy data itself. DNNs, on the other hand, are capable of extracting these features automatically.

\begin{figure*}
		\includegraphics[width=\textwidth]{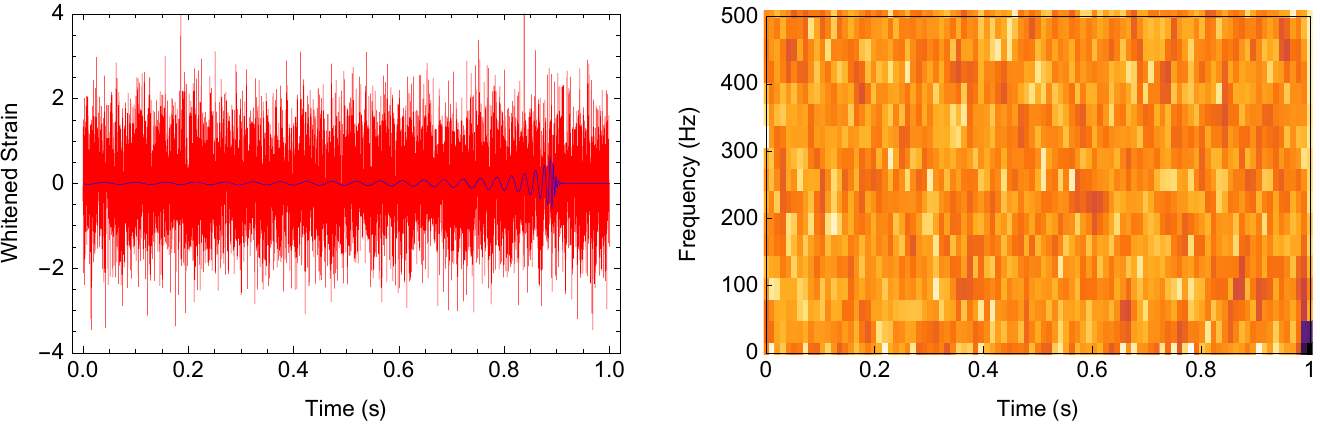}
	\caption{\textbf{Sample of input data}. The red time-series is an example of the input to our DNN algorithm. It contains a BBH GW signal (blue) which was whitened with aLIGO's design sensitivity and superimposed in noisy data with SNR = 7.5 (peak power of this signal is 0.36 times the power of background noise). The component masses of the merging BHs are $57\Msun$ and $33\Msun$. The corresponding spectrogram on the right shows that the GW signal on the left is not visible, and thus cannot be detected by an algorithm trained for image recognition. Nevertheless, our DNN detects the presence of this signal directly from the (red) time-series input with over 99\% sensitivity and reconstructs the source's parameters with a mean relative error of about 10\%. }
	\label{fig:spectrogram}
\end{figure*}

Deep learning has been previously applied only for the classification of glitches with spectrogram images as inputs to CNNs~\cite{GravitySpy,GravitySpy2,Geo:2017b} and unsupervised clustering of transients~\cite{Geo:2017b}, in the context of aLIGO. Using images as inputs is advantageous for two reasons: (i) there are well established architectures of 2D CNNs which have been shown to work (GoogLeNet~\cite{GoogLeNet}, VGG~\cite{VGG}, ResNet~\cite{ResNet}) and (ii) pre-trained weights are available for them, which can significantly speed up the training process via transfer learning while also providing higher accuracy even for small datasets~\cite{Geo:2017b}. However, our experiments showed that this approach would not be optimal for detection or parameter estimation since many signals having low signal-to-noise ratio (SNR~\footnote{Note that we are using the standard definition of optimal matched-filtering SNR, as described in~\cite{saton}. This SNR is on average proportional to $12.9\pm1.4$ times the ratio of the amplitude of the signal to the standard deviation of the noise for our test set.}) are not visible in spectrograms, as shown in Fig.~\ref{fig:spectrogram}. 

\noindent Theoretically, all the information about the signal is encoded within the time-series, whereas spectrograms are lossy non-invertible representations of the original data. Although 2D CNNs are commonly used, especially for image-related tasks, we found that by directly feeding raw time-series data as inputs to certain types of CNNs, one can obtain much higher sensitivities at low SNR, significantly lower error rates in parameter estimation, and faster analysis speeds. This automated feature learning allows the algorithm to develop more optimal strategies of signal processing than when given hand-extracted information such as spectrograms. There has only been a few attempts at signal processing using CNNs with raw time-series data in general and only for single parameter estimation~\cite{ConvNet-Radio,ConvNet-TimeSeries}. 

In this work, we demonstrate, for the first time, that DNNs can be used for both signal detection and multiple-parameter estimation directly from highly noisy time-series data, once trained with templates of the expected signals, and that dilated CNNs outperform traditional machine learning algorithms, and reach accuracies comparable to matched-filtering methods. We also show that our algorithm is far more computationally efficient than matched-filtering. Instead of repeatedly performing overlap computations against all templates of known signals, our CNN builds a deep \textit{non-linear} hierarchical structure of nested convolutions, with small kernels, that determines the parameters in a single evaluation. Moreover, the DNNs act as an efficient compression mechanism by learning patterns and encoding all the relevant information in their weights, analogous to a reduced-order model~\cite{pu:2016PhRvD}, which is significantly smaller than the size of the training templates. Therefore, the DNNs automatically perform an internal optimization of the search algorithm and can also interpolate, or even extrapolate, to new signals not included in the template bank (unlike matched-filtering).

Note that matched-filtering is equivalent to a single convolution layer of a neural network, with very long kernels corresponding to all the signals in a template bank. Therefore, our algorithm can be viewed as an extension of matched-filtering, which performs template matching against a small set of short duration templates, and aggregates this information in the deeper layers to effectively model the full range of long-duration signals.


\section{Method}
\label{meth}

Our goal is to show that \texttt{Deep Filtering} is a powerful tool for GW data analysis. We do this by demonstrating that a system of two DNNs can \textit{detect} and \textit{characterize} GW signals embedded in highly noisy time-series data.

As a proof of concept, we focus on GWs from BBH mergers, which are expected to dominate the number of GW detections with ground-based GW detectors~\cite{bel:2016Na,Belc:2015BB,bbhswithligo:2016}. In future work, we will extend this method to signals produced by other events by adding more neurons in the final layer and training with larger datasets.

We chose to divide the problem into two separate parts, each assigned to a different DNN. The first network, henceforth known as the ``classifier'', will detect the presence of a signal in the input, and will provide a confidence level for the detection. The classes chosen for now are ``True'' or ``False'' depending on whether or not a signal from a BBH merger is present in the input. The second network, which we call the ``predictor'', will estimate the parameters of the source of the signal (in this case, the component masses of the BBH). The predictor is triggered when the classifier identifies a signal with a high probability. 

We partitioned the system in this manner so that, in the future, more classes of GW transients~\cite{2016PhRvD..93l4062H,ETL:2012CQGra,2014CQGra..31a5005M}, may be added to the classifier, and separate predictors can be made for each type of signal. Moreover, categories for various types of anomalous sources of noise, like glitches and blips~\cite{GravitySpy,corn:2015CQGra}, can also be added to the classifier~\cite{Geo:2017b}.

\subsection*{Assumptions}
\label{assume}

For this initial study, we have assumed the signals are optimally oriented with respect to the detectors, and that the individual spins and orbital eccentricities are zero. This reduces our parameter space to two dimensions, namely, the individual masses of the BBH systems, which we have restricted to lie between $5\Msun$ and $75\Msun$. Furthermore, we have constrained the inputs to have a duration of 1 second, and a sampling rate of 8192Hz throughout this analysis, which is more than sufficient for the events we are considering.  Note that the
classifier will be applied to the continuous data stream by using a sliding window of width 1 second.

\begin{figure}
	\centering
	\includegraphics[width=0.465\textwidth]{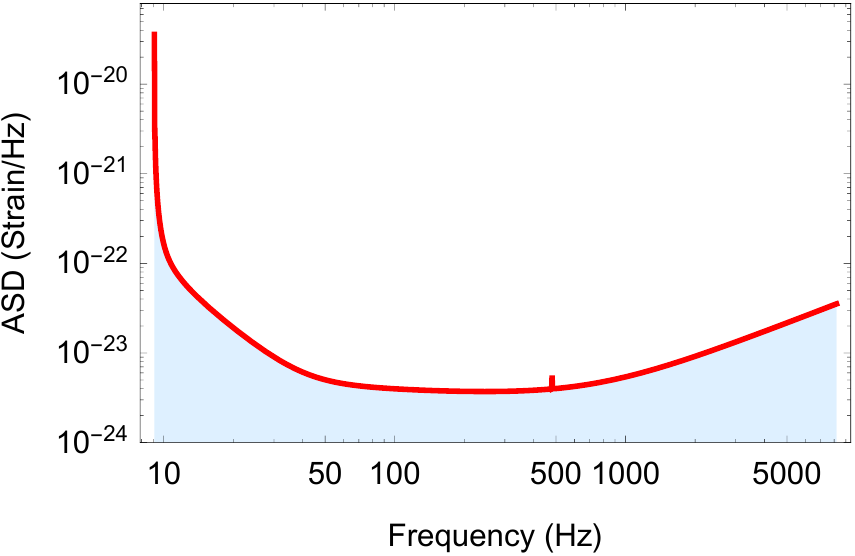}
	\caption{Throughout this analysis, we have used the Zero Detuned High Power sensitivity configuration for aLIGO~\cite{ZDHP:2010} to simulate the colored noise in the detectors.}
	\label{ASD}
\end{figure}

Ideally, the inputs to our DNNs will be the unprocessed time-series of strains measured by the GW detectors. Throughout this analysis, however, we have whitened the signals using aLIGO's Power Spectral Density (PSD) at the ``Zero-detuned High Power'' design sensitivity~\cite{ZDHP:2010} shown in Figure~\ref{ASD}. We have also ignored glitches, blips, and other transient sources of detector noise for now. This is in line with previous studies, which have first showcased a machine learning algorithm for LIGO data analysis using simulated noise~\cite{jade:2015CQGra,bambiann:2015PhRvD}, and then followed up by an independent study where the algorithm is tested using real aLIGO noise~\cite{jade1:2016}. Our analysis, using real aLIGO data, will be presented in a separate publication.

\subsection*{Obtaining Data}

Supervised deep learning algorithms are far more effective when trained with very large datasets. Obtaining high quality training data has been a difficult and cumbersome task in most applications of DNNs, such as object recognition in images, speech and text processing, etc. Fortunately, we do not face this issue, since we can take advantage of scientific simulations to produce the necessary data for training. 

Over the last decade, sophisticated techniques have been developed to perform accurate 3-dimensional NR simulations of merging BHs~\cite{Mroue:2013,ETL:2012CQGra} on HPC facilities. For the analysis at hand, we use Effective-One-Body (EOB) waveforms that describe GWs emitted by quasi-circular, non-spinning BBHs~\cite{Tara:2014}. We extracted the final 1 second window of each template for our analysis.

\begin{figure}
	\hspace{-.2in}	\includegraphics[width=.48\textwidth]{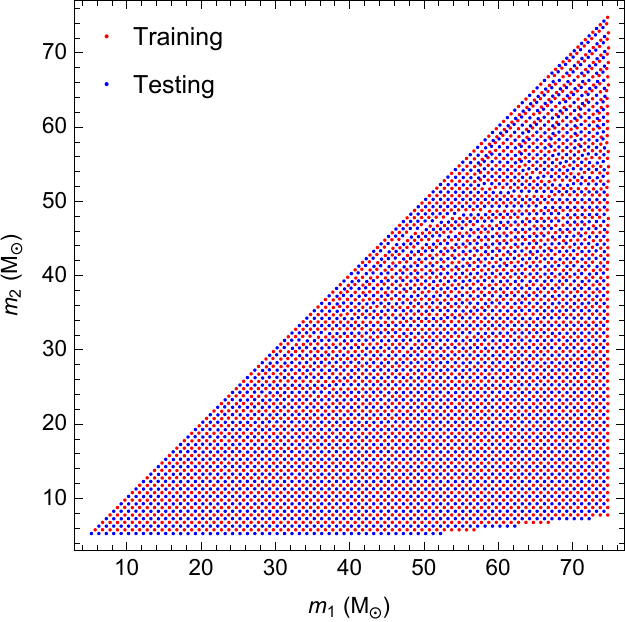}\hspace{6em}
	\caption{\textbf{Distribution of data.} The figure shows the distribution of component masses of BBHs for the training and testing datasets. The mass-ratios were confined between 1 and 10, which accounts for the missing points in the lower right corner. We choose this mass-ratio range because the state-of-the-art EOB model we have used to create the datasets has only been validated for these mass-ratio values. Each point represents a quasi-circular, non-spinning GW signal of 1 second duration, sampled at 8192 Hz, which is whitened with aLIGO's expected noise spectrum at design sensitivity. These waveforms were normalized and translated randomly in time. Thereafter, multiple batches of noise at each SNR were added to produce training and testing datasets.}
	\label{fig:m1m2}
\end{figure}

We have split the data into separate sets for training and testing. For the training dataset, the BBHs component masses are in the range $5\Msun$ to $75\Msun$ in steps of $1\Msun$. The testing dataset has intermediate component masses, i.e., masses separated from values in the training dataset by $0.5\Msun$. By not having overlapping values in the training and testing sets, one can ensure that the network is not overfitting, i.e., memorizing only the inputs shown to it without learning to generalize to new inputs. The distribution of component masses, and a template from the training and testing sets, is shown in Fig.~\ref{fig:m1m2}.

\noindent Subsequently, we shifted the location of the peak of each signal randomly within an interval of 0.2 seconds in both the training and testing sets to make the DNNs more robust with respect to time translations. Next, we superimposed different realizations of Gaussian white noise on top of the signals over multiple iterations, thus amplifying the size of the datasets. The power of the noise was adjusted according to the desired SNR for each training session. We then standardized the inputs to have zero mean and unit variance to make the training process easier~\cite{LeCun98}.

The final training sets at each SNR were produced from  \(\sim 2500\) templates of GWs from BBH mergers by adding multiple batches of noise and shifting in time. It is also a standard practice to use a validation set to monitor the performance on unseen data during training in order to prevent overfitting. The validation and testing sets at each SNR were generated from a different set of \(\sim 2500\) templates by superimposing different noise realizations. 

\subsection*{Designing Neural Networks}

We used very similar DNN architectures for both the classifier and predictor, which demonstrates the versatility of this method. The only difference was the addition of a softmax layer to the classifier to obtain probability estimates as the outputs. Our strategy was to first train the predictor on the datasets labeled with the BBH masses, and then transfer the weights of this pre-trained network to initialize the classifier and then train it on datasets with 50\% random noise. This transfer learning process reduced the training time required for the classifier, while also slightly improving its accuracy at low SNR.

We designed simple DNNs from the ground up. Overall, we tested around 80 configurations of DNNs ranging from 1 to 4 convolutional layers and 1 to 3 fully connected layers (also called linear layers) similar to~\cite{lecun98-cnn}, but modified for time-series inputs. Among these, we discovered that a design for the classifier with 3 convolutional layers followed by 2 fully connected layers yielded good results with fastest inference speed for the datasets that we are considering. We tried adding a few recent developments such as batch normalization~\cite{BatchNormalization} and dropout~\cite{Dropout} layers. However, we did not use them in our final design as they did not provide significant improvements for the simple problem we are considering. The addition of noise to the signals during the training process serves as a form of regularization in itself. Many of the layers have parameters, commonly known as hyperparameters, which we had to tune manually via a randomized trial-and-error procedure. 

\begin{figure}
	\centering
	\includegraphics[width=0.38\textwidth]{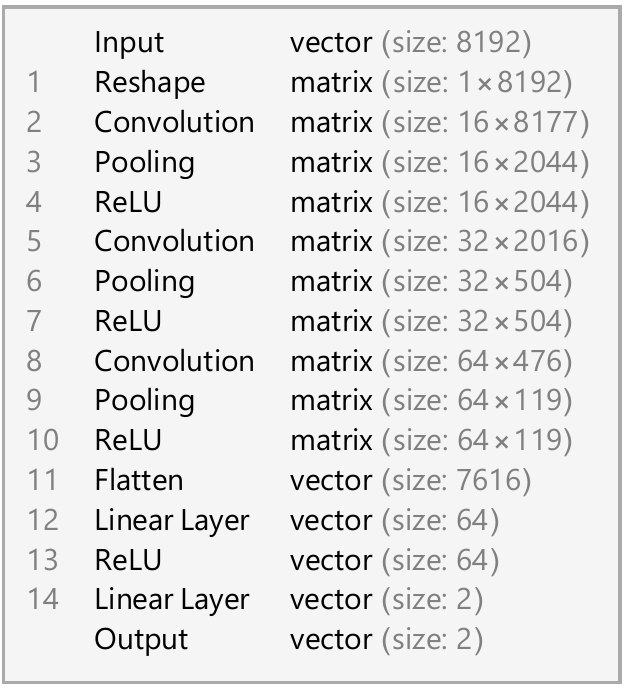}
	\caption{\textbf{Architecture of deep neural network}. This is the deep dilated 1D CNN, modified to take time-series inputs, that we designed for prediction which outputs two real-valued numbers for the two component masses of the BBH system. For classification we simply added a softmax layer after the 14th layer to obtain the probabilities for two classes, i.e., ``True" or ``False". The input is the time-series sampled at 8192Hz and the output is either the probability of each class or the value of each parameter. Note that the number of neurons in layer 14 can be increased to add more categories for classification or more parameters for prediction. The size of this net is about 2\textit{MB}.}
	\label{fig:Classifier}
\end{figure}

\begin{figure}
	\centering
	\includegraphics[width=0.37\textwidth]{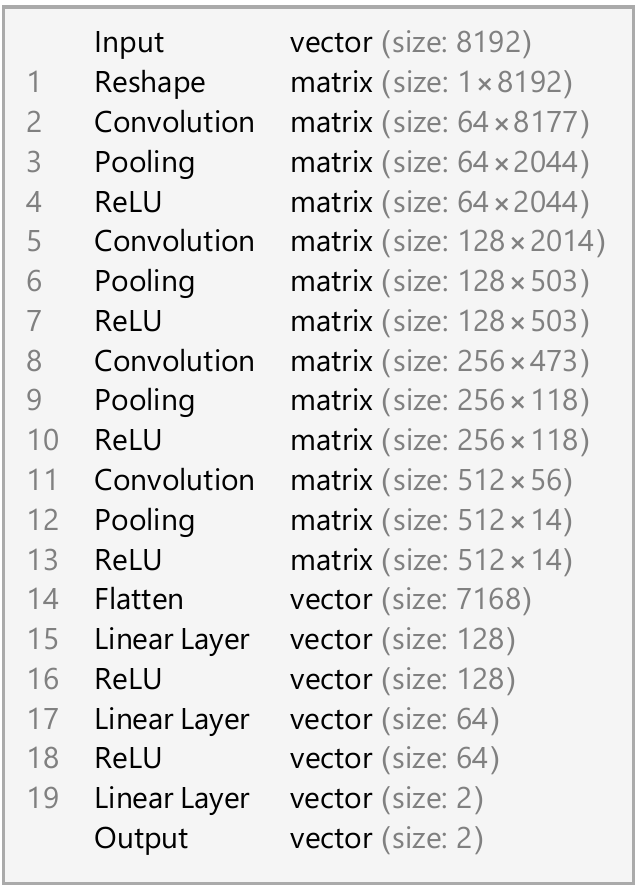}
	\caption{\textbf{Architecture of deeper neural network}. This is the deeper version of the CNN, modified to take time-series inputs, that we designed for parameter estimation. The input is the time-series sampled at 8192Hz and the output is the predicted value of each parameter. This can be converted to a classifier by adding a softmax layer after layer 19 to obtain the probability for a detection. Note that the number of neurons in layer 19 can be increased to add more categories for classification or more parameters for prediction. The 2 neurons in the final layer outputs the 2 parameters corresponding to the individual masses of BBHs. The size of this net is approximately 23\textit{MB}.}
	\label{fig:Predictor}
\end{figure}

\noindent Depth is a hyperparameter which determines the number of filters in each convolutional layer. Our choices for depth in the consecutive layers were 16, 32, and 64 respectively. We used kernel sizes of 16, 8, and 8 for the convolutional layers and 4 for all the (max) pooling layers. Stride, which specifies the shift between the receptive fields of adjacent neurons, was chosen to be 1 for all the convolution layers and 4 for all the pooling layers. Dilation determines the overall size of each receptive field, which could be larger than the kernel size by having gaps in between. Here, it is a measure of the temporal extend of the convolutions. We observed that using dilation of 4 in the final two convolution layers improved the performance. The final layout of our classifier DNN is shown in Fig.~\ref{fig:Classifier}. 

Deeper networks are expected to provide further improvements in accuracy although at the cost of slower evaluation speed. To show this, we also designed a deeper net, shown in Fig.~\ref{fig:Predictor}, with 4 convolution layers and 3 fully connected layers that had comparable sensitivity for detection and significantly better performance for parameter estimation. Although this design performed slightly better, it was a factor of 5 slower on a GPU for evaluation.  This net had convolution layers having kernel sizes were 16, 16, 16, and 32 with dilations 1, 2, 2, and 2 respectively. The pooling layers all had kernel size 4 and stride 4.

A loss function (cost function) is required to compute the error after each iteration by measuring how close the outputs are with respect to the target values. We designed a mean absolute relative error loss function for the predictor. For classification, we used the standard cross-entropy loss function.

\subsection*{Training Strategy}

We spent significant effort on hyperparameter optimization, to design architectures of the CNNs by trial and error. First, we used Gaussian white noise without whitening the signals i.e., a flat PSD, to determine the optimal architectures of the DNNs. We found that this design was also optimal for signals whitened with the Zero-Detuned PSD of aLIGO. This indicates that the same architecture will perform well on wide variety of PSDs. Once we chose the best performing DNNs, we trained it for about a total of 10 hours. We relied on the neural network functionality in the Wolfram Language, \textit{Mathematica}, based internally on the open-source MXNet framework~\cite{MXNet}, which utilizes the CUDA deep learning library (cuDNN)~\cite{cuDNN} for acceleration with NVIDIA GPUs. We used the ADAM~\cite{ADAM} method as our learning algorithm. 

\noindent During this process, we developed a new strategy to improve the performance and reduce training times of the DNNs. By starting off training the predictor on inputs having high SNR ($\ge 100$) and then gradually increasing the noise in each subsequent training session until a final SNR distribution randomly sampled in the range 5 to 15, we observed that the performance can be quickly maximized for low SNR, while remaining accurate for signals with very high SNR. For instance, we obtained about 11\% error when trained using this scheme with gradually decreasing SNR and only about 21\% mean error at parameter estimation on the test set when directly trained on the same range of SNR (5-15).  Furthermore, we found that the classifier performs significantly better (with an increase from 96\% to 99\% accuracy on one of our test sets) when its initial weights are transfered from the fully trained predictor, i.e., the classifier was created by simply adding a softmax layer to the trained predictor and then trained on the dataset of signals and noise. We expect these techniques would be useful for training neural networks, in general, with noisy data. 


\section{Results} 
\label{result}

\begin{figure}
	
	\hspace{-.25in}	\includegraphics[width=.505\textwidth]{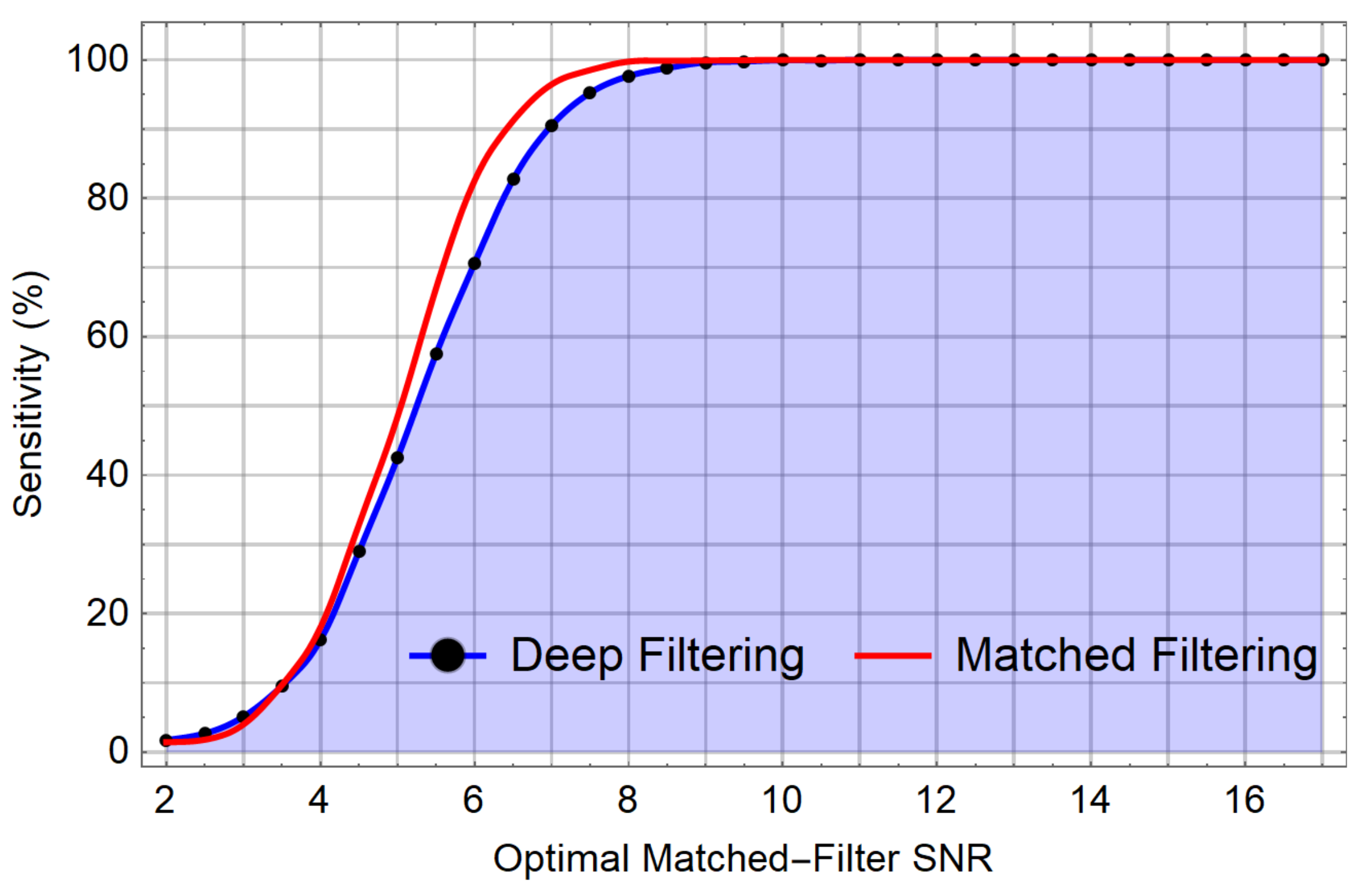}
	
	\caption{\textbf{Sensitivity of detection with smaller net.} This is the sensitivity (fraction of signals detected) of the shallower classifier as a function of SNR on the test set. Note that the sensitivity was measured with the same classifier after training once over the entire range of SNR, i.e., without specifically re-training it for each SNR. This curve saturates at sensitivity of 100\% for SNR $\ge 10$, i.e, signals with \(\textrm{SNR} \geq 10\) are always detected. The single detector false alarm rate was tuned to be about 0.5\% for this classifier. Note that the optimal matched-filter SNR is on average proportional to $12.9\pm1.4$ times the ratio of the amplitude of the signal to the standard deviation of the noise for our test set. This implies that Deep Filtering is capable of detecting signals significantly weaker than the background noise.}
	\label{fig:Sensitivity}
\end{figure}

\begin{figure}
	
	\hspace{-.25in}	\includegraphics[width=.51\textwidth]{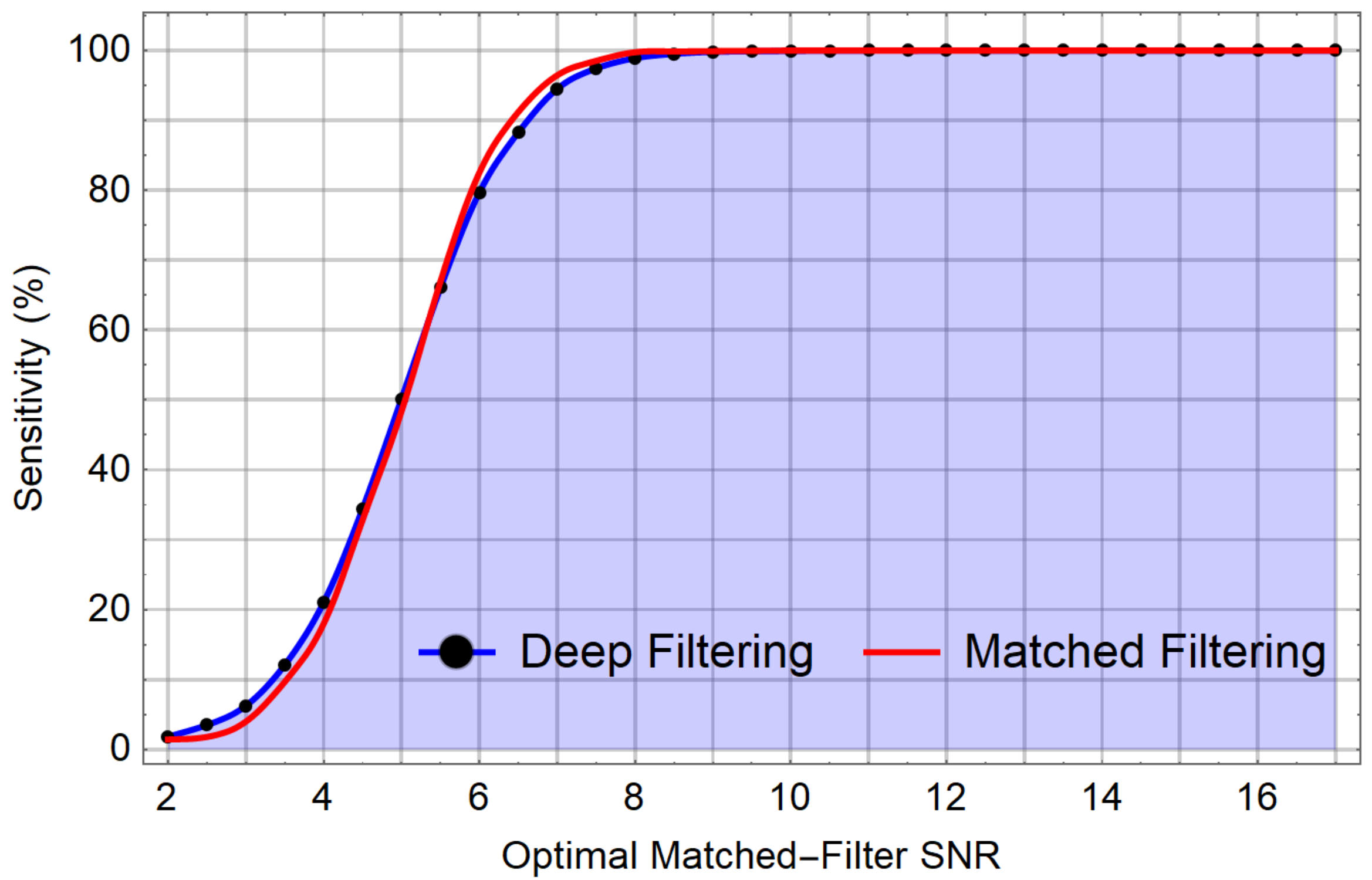}
	
	\caption{\textbf{Sensitivity of detection with deeper net.} This is the sensitivity of the deeper classifier as a function of SNR on the test set. Note that this sensitivity was also measured with the same classifier after training once over the entire range of SNR, i.e., without specifically re-training it for each SNR. This curve saturates at sensitivity of 100\% for SNR $\ge 9$, i.e, signals with \(\textrm{SNR} \geq 9\) are always detected. The single detector false alarm rate was tuned to be approximately 0.5\% for this classifier. }
	\label{fig:SensitivityNew}
\end{figure}

We trained our classifier to achieve 100\% sensitivity for signals with SNR $\geq 10$ and a single detector false alarm rate less than 0.6\%. Note that the false alarm rate of \texttt{Deep Filtering} can be significantly decreased by combining classifications on multiple detector inputs and by computing the overlap of the template predicted by \texttt{Deep Filtering} with the input to confirm each detection. The sensitivity of this classifier as a function of SNR is shown in Fig.~\ref{fig:Sensitivity}. The deeper classifier obtained slightly better sensitivity as shown in Fig.~\ref{fig:SensitivityNew}

\begin{figure*}
	\centerline{
		\includegraphics[width=.47\textwidth]{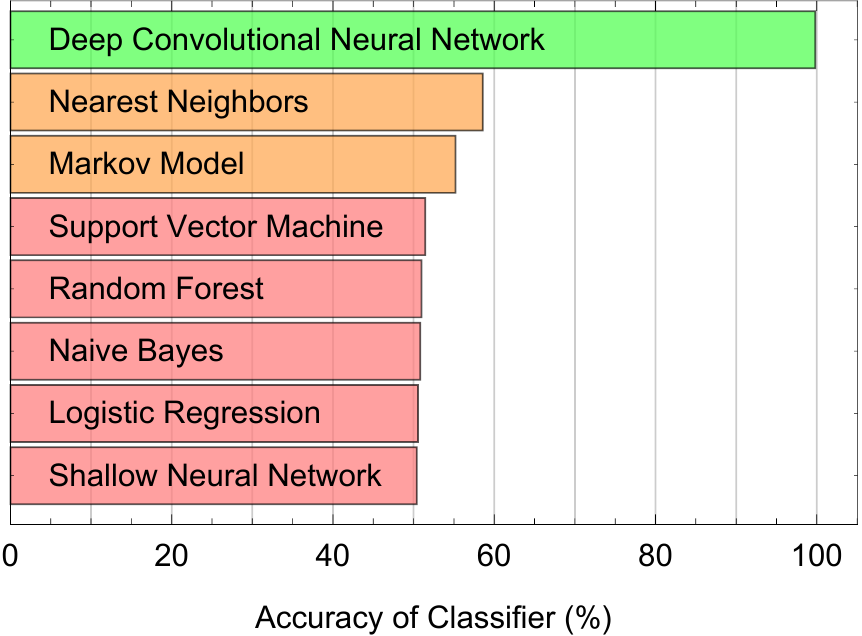}\hspace{2.1em}
		\includegraphics[width=.47\textwidth]{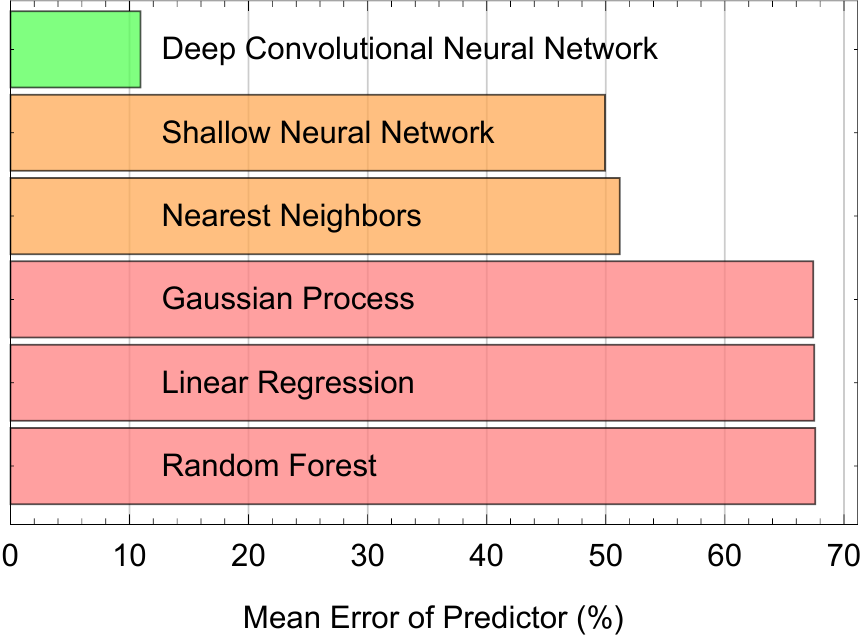}
	}

	\caption{\textbf{Comparison with other methods.} Left panel: This is the accuracy of different machine learning methods for detection after training each with roughly 8000 elements, half of which contained noisy signals with a fixed peak power, less than the background noise, and constant total mass, with the other half being pure noise with unit standard deviation.
		An accuracy of 50\% can be obtained by randomly guessing. Right panel: This is the mean relative error obtained by various machine learning algorithms for predicting a single parameter, i.e., mass-ratio, using a training set containing about 8000 signals with fixed amplitude = 0.6 added to white noise with unit standard deviation. Note that scaling these methods to predict multiple parameters is often difficult, whereas it simply involves adding neurons to the final layer of neural networks.}
	\label{fig:ML}
\end{figure*}

\noindent For comparison, we trained standard implementations of all commonly used machine learning classifiers--- Random Forest, Support Vector Machine, k-Nearest Neighbors, Hidden Markov Model, Shallow Neural Networks, Naive Bayes, and Logistic Regression --- along with the DNNs on a simpler training set of 8000 elements for fixed total mass and peak signal amplitude. Unlike DNNs, none of these algorithms were able to directly handle raw noisy data even for this simple problem as shown in Fig.~\ref{fig:ML}.

\begin{figure}

	\hspace{-.2in}	\includegraphics[width=.5\textwidth]{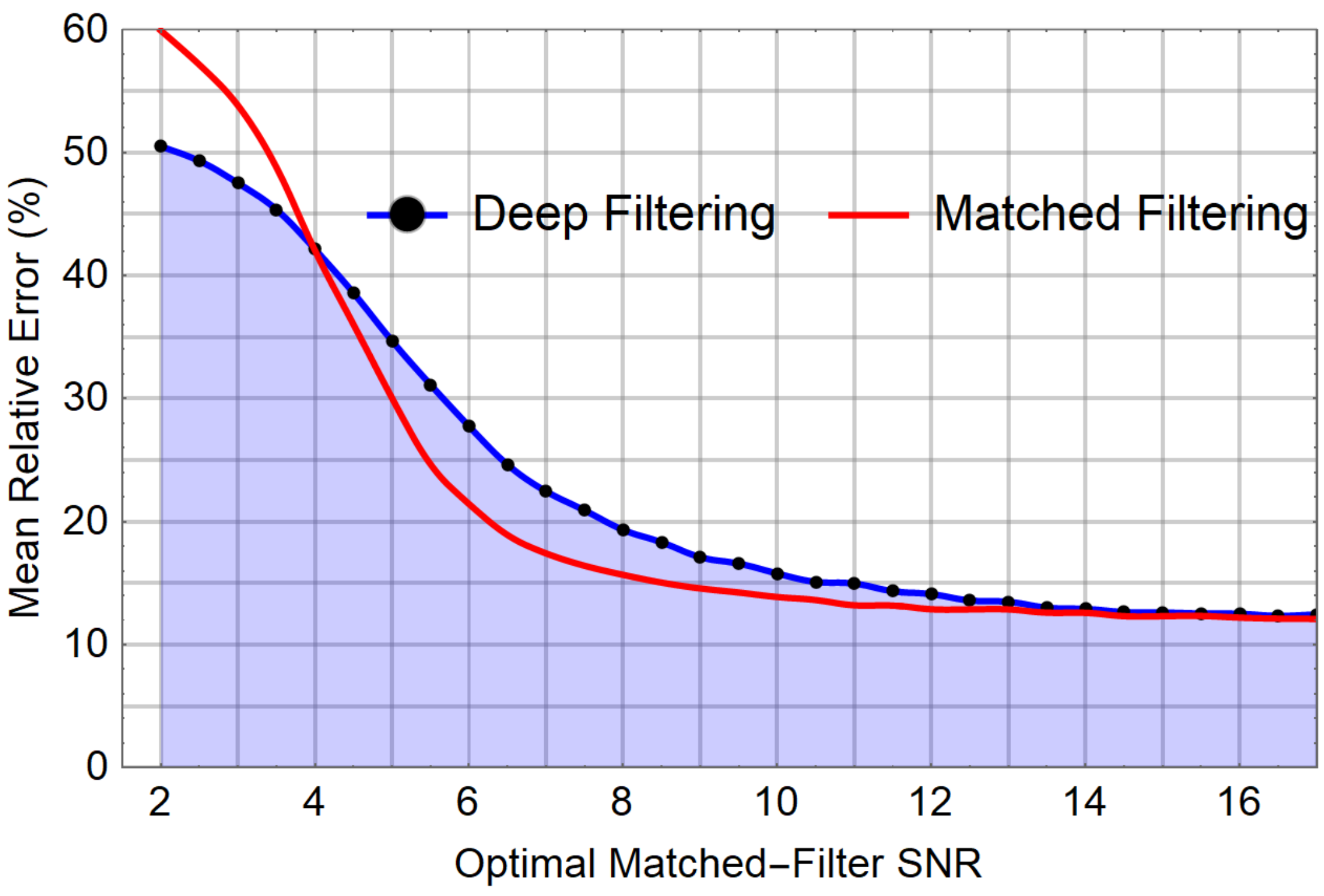}\hspace{1.7em}

	\caption{\textbf{Error in parameter estimation with smaller net}. This shows the mean percentage error of estimated masses on our testing sets at each SNR using the predictor DNN with 3 convolution layers. The DNN was trained only once over the range of SNR and was then tested at different SNR, without re-training. Note that a mean relative error less than 20\% was obtained for SNR $\ge8$ . At high SNR, the mean error saturates at around 11\%. See Fig.~\ref{fig:ErrPredict2} for the results with the deeper version of the predictor.
}
	\label{fig:ErrPredict}
\end{figure}

\begin{figure}
	
	\hspace{-.2in}	\includegraphics[width=.5\textwidth]{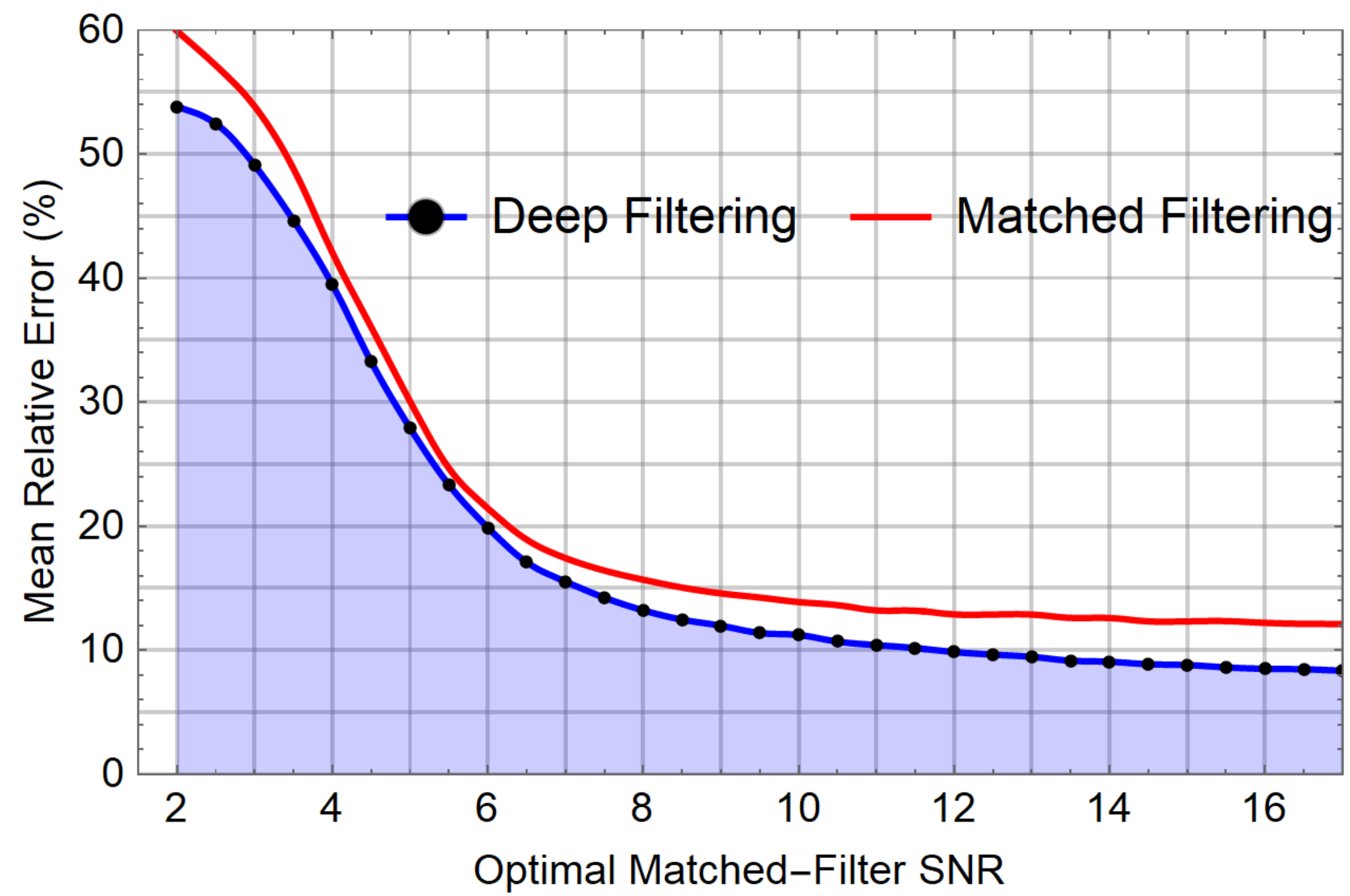}
	
	\caption{\textbf{Error in parameter estimation with deeper net}. This shows the mean percentage error of estimated masses on our testing sets at each SNR using the deeper CNN with 4 convolution layers. Note that a mean relative error less than 15\% was obtained for SNR $\ge 7$ . At high SNR, the mean error saturates at around 7\%. Note that we were able to optimize the predictor to have less than 3\% error for very high SNR ($\ge50$), which demonstrates the ability of \texttt{Deep Filtering} to learn patterns connecting the templates and effectively interpolate to intermediate points in parameter space.
	}
	\label{fig:ErrPredict2}
\end{figure}

\noindent Our predictor was able to successfully measure the component masses given noisy GWs, that were not used for training, with an error of the same order as the spacing between templates for SNR $\ge 13$. The deeper predictor consistently outperformed matched-filtering. At very large SNR, over 50, we could train both the predictors to have relative error less than 5\%, whereas the error with matched-filtering using the same templates was always greater than 11\% with the given template bank. This means that, unlike matched-filtering, our algorithm is able to automatically perform interpolation between the known templates to predict intermediate values. The variation in relative error against SNR for each architecture of the DNNs is shown in Fig.~\ref{fig:ErrPredict} and Fig.~\ref{fig:ErrPredict2}. The largest relative errors were concentrated at lower masses, because a small variation in predicted masses led to larger relative errors in this region.

We can estimate the distribution of errors and uncertainties \textit{empirically} at each region of the parameter-space. We observed that the errors closely follow Gaussian normal distributions for each input for SNR ($\ge 9$), allowing easier characterization of uncertainties. Once we obtain initial estimates for the parameters via \texttt{Deep Filtering}, traditional techniques may be rapidly applied using only a few templates near these predictions to cross-validate our detection and parameter estimates and to measure uncertainties. There are also emerging techniques to estimate quantify in the predictions of CNNs~\cite{images:2017}, which may be applied to this method.

\begin{figure}
	
	\hspace{-.25in}	\includegraphics[width=.475\textwidth]{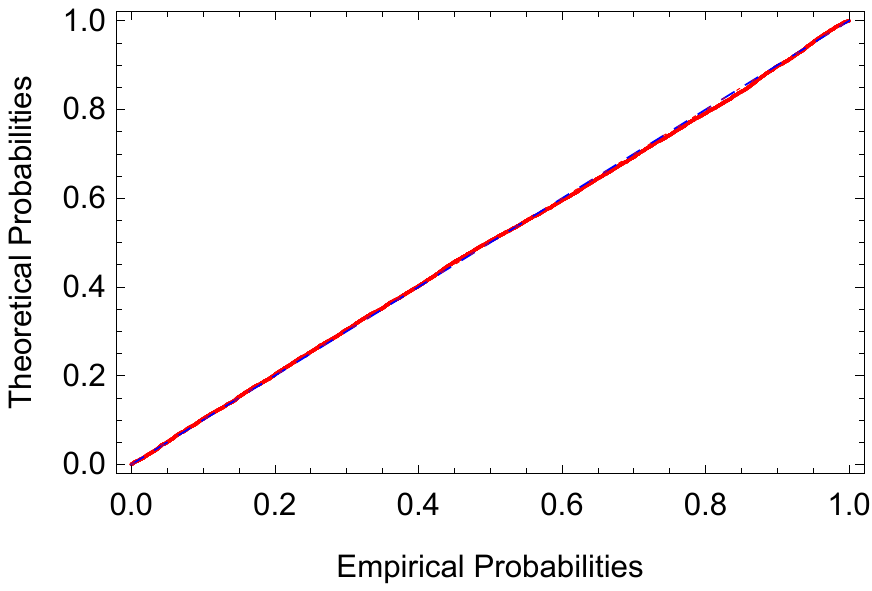}
	\caption{\textbf{P-P plot of errors in parameter estimation} This is a P-P (probability) plot of the distribution of errors in predicting $m_1$ for test  parameters $m_1=57M_{\odot}$ and $m_2=33M_{\odot}$, superimposed with different realizations of noise at SNR = 9. The best-fit is a Gaussian normal distribution with mean $=1.5M_{\odot}$ and standard deviation $=4.1M_{\odot}$. The errors have similar Gaussian distributions in other regions of the parameter-space as well. }
	\label{fig:PPplot}
\end{figure}

\noindent After testing common machine learning techniques including Linear Regression, k-Nearest Neighbors, Shallow Neural Networks, Gaussian Process Regression, and Random Forest on the simpler problem with fixed total mass, we observed that, unlike DNNs, they could not predict even a single parameter (mass-ratio at fixed total mass) accurately, as evident from Fig.~\ref{fig:ML}, when trained directly on time-series data.

\begin{figure*}
	\centerline{\hspace{-.045in}
		\includegraphics[width=.495\textwidth]{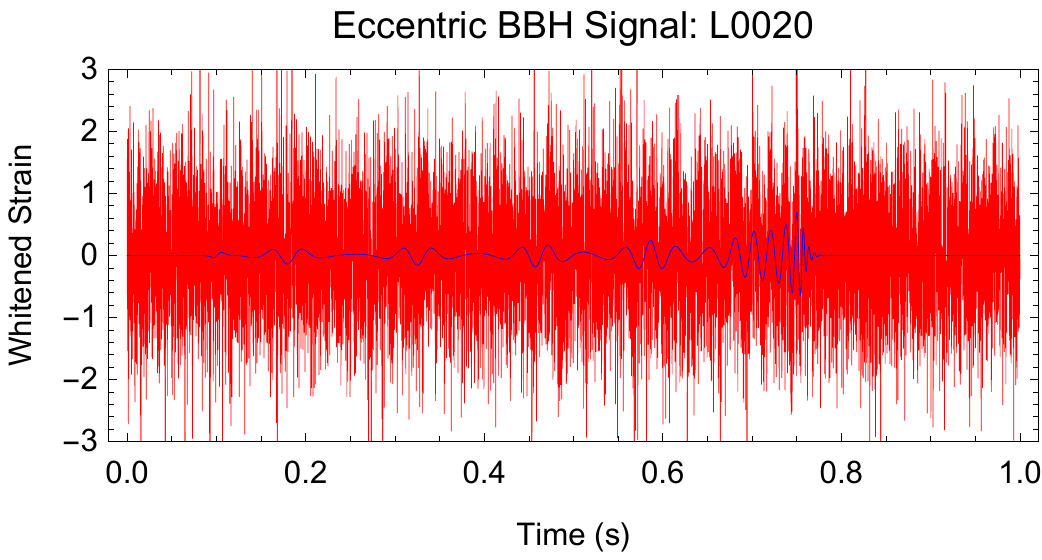}\hspace{.5em}
		\includegraphics[width=.495\textwidth]{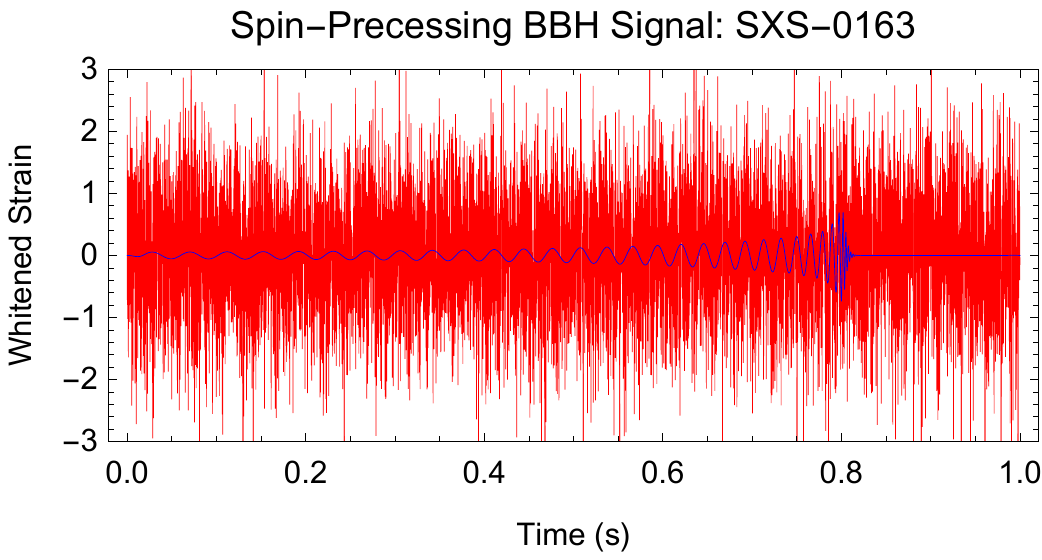}
	}
	\caption{\textbf{New types of signals}. Left panel: This waveform was obtained from one of our NR simulations of eccentric BBH merger that has mass-ratio 5.5, total mass about $90\Msun$, and an initial eccentricity $e_0=0.2$ when it enters the aLIGO band. Our \texttt{Deep Filtering} pipeline successfully detected this signal, even when the total mass was scaled between $50\Msun$ and $90\Msun$, with 100\% sensitivity (for SNR $\ge 10$) and predicted the component masses with a mean relative error $\le30\%$ for SNR $\ge 12$. 
		Right panel: One of the spin-precessing waveforms obtained from the NR simulations in the SXS catalog with component masses equal to $25\Msun$ each. The individual spins are each 0.6 and oriented in un-aligned directions. Our DNNs also successfully detected this signal, even when the total mass was scaled between $40\Msun$ and $100\Msun$, with 100\% sensitivity for SNR $\ge 10$ and predicted the component masses with a mean relative error $\le20\%$ for SNR $\ge 12$. 
	}
	\label{fig:ecc_pre}
\end{figure*}

\noindent Having trained our DNNs to detect and characterize quasi-circular, non-spinning BBH signals, we assessed their capabilities to identify new classes of GW signals, beyond our original training and testing sets. We used two distinct types of signals that were \textit{not} considered during the training stage, namely: (i) moderately eccentric NR simulations (approximate eccentricity of 0.1 when entering aLIGO band), that we recently generated with the open-source, Einstein Toolkit~\cite{ETL:2012CQGra} using the Blue Waters petascale supercomputer; and (ii) NR waveforms from the SXS catalog~\cite{chu:2016CQG} that describe spin-precessing, quasi-circular BBHs---each BH having spin $\ge 0.5$ oriented in random directions~\cite{chu:2016CQG}. Sample waveforms of these GW classes as shown in Fig.~\ref{fig:ecc_pre}. Since these NR simulations scale trivially with mass, we enlarged the data by rescaling the signals to have different total masses. Thereafter, we whitened them and added different realizations of noise, in the same manner as before, to produce test sets.

We have found that both the classifiers detected all these signals with nearly the same rate as the original test set, with 100\% sensitivity for SNR $\ge 10$. Remarkably, the predictor quantified the component masses of our eccentric simulations for SNR $\ge 12$ with a mean relative error less than 20\%  for mass-ratios \(q=\{1,\,2,\,3,\,4\}\), and less than 30\% for \(q=5.5\) respectively. For the spin-precessing systems we tested, with SNR $\ge 12$, the mean error in predicting the masses was less than 20\% for \(q=\{1,\,3\}\), respectively.

These findings are very encouraging, since recent analyses have made evident that existing aLIGO algorithms are not capable of accurately detecting or reconstructing the parameters of eccentric signals~\cite{Huerta:2017a,Huerta:2014,Huerta:2013a}, and do not cover spin-precessing systems~\cite{2016CQGra..33u5004U}.  This ability to generalize to new categories of signals, without being shown any examples, means that DNN-based pipelines can increase the depth of existing GW detection algorithms without incurring in any additional computational expense. 

\begin{figure}
	\centering
	\includegraphics[width=0.45\textwidth]{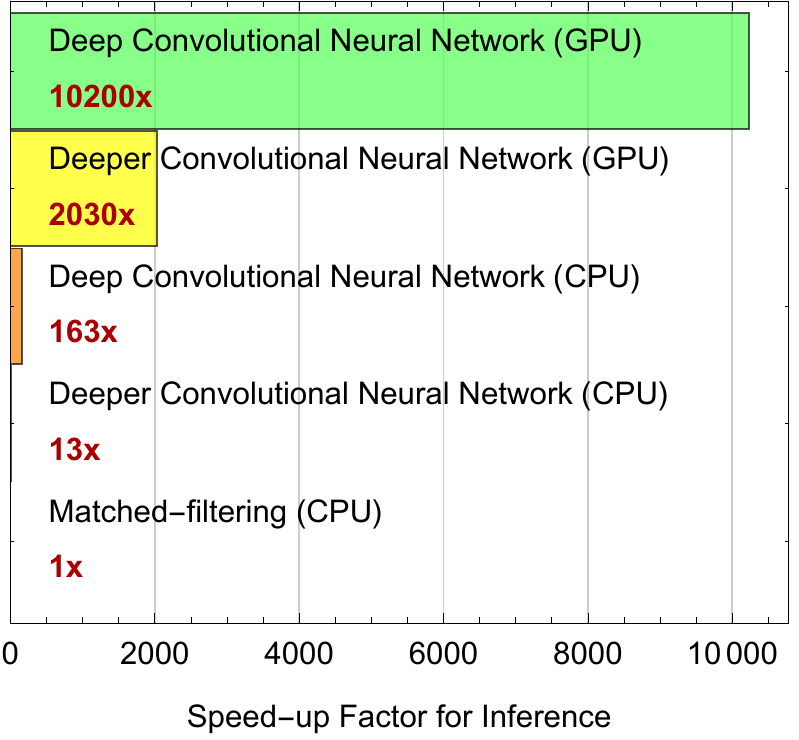}
	\caption{\textbf{Speed-up of analysis}. The DNN-based pipeline is many orders of magnitude faster compared to matched-filtering (cross-correlation or convolution) against the same template bank of waveforms (tested on batches of inputs using both cores of an Intel Core i7-6500U CPU and an inexpensive NVIDIA GeForce GTX 1080 GPU for a fairer comparison). Note that the evaluation time of a DNN is constant regardless of the size of training data, whereas the time taken for matched-filtering is proportional to the number of templates being considered, i.e., exponentially proportional to the number of parameters. Therefore, the speed-up of \texttt{Deep Filtering} would be higher in practice, especially when considering larger template banks over a higher dimensional parameter space.}
	\label{fig:speed}
\end{figure}

\noindent Furthermore, our simple classifier and predictor are only 2\textit{MB} in size each, yet they achieve excellent results. The average time taken for evaluating them per input of 1 second duration is approximately 6.7 milliseconds, and 106 microseconds using a single CPU and GPU respectively. The deeper predictor net, which is about 23\textit{MB}, achieves slightly better accuracy at parameter estimation but takes about 85 milliseconds for evaluation on the CPU and 535 microseconds on the GPU, which is still orders of magnitude faster than real-time. Note that the current deep learning frameworks are not well optimized for CPU evaluation. For comparison, we estimated an evaluation time of 1.1 seconds for time-domain matched-filtering on the same CPU (using 2-cores) with the same template bank of clean signals used for training, the results are shown in Fig.~\ref{fig:speed}. This extremely fast inference rate indicates that real-time analysis can be carried out with a single CPU or GPU, even with DNNs that are significantly larger and trained over a much larger template banks of millions of signals. For example, a state-of-the-art CNN for image recognition~\cite{ImageIdentify,BatchNorm} has hundreds of layers (61\textit{MB} in size) and is trained with over millions of examples to recognize thousands of different categories of objects. This CNN can process significantly larger inputs, each having dimensions $224\times224\times3$, using a single GPU with a mean time of 6.5 milliseconds per input. Note that these CNNs can be trained on millions of inputs in a few hours using parallel GPUs~\cite{FastTrain}.

For applying the \texttt{Deep Filtering} method to a multi-detector scenario, we simply need to apply our nets pre-trained for single detector inference separately to each detector and check for coincident detections with similar parameter estimates. Enforcing coincident detections would decrease our false alarm probability, from about $0.59\%$ to about $0.003\%$. Once the \texttt{Deep Filtering} pipeline detects a signal then traditional matched-filtering may be applied with a select few templates around the estimated parameters to cross-validate the event and estimate confidence measure. Since only a few templates need to be used with this strategy, existing challenges to extend matched-filtering for higher dimensional GW searches may thus be overcome, allowing real-time analysis with minimal computational resources.

\section{Discussion}
\label{disc}

The results we obtained with our prototype DNNs exceeded our expectations with high detection rate and low prediction errors even for signals with very low SNR. Initially, we had trained a DNN to predict only the mass-ratios at a fixed total mass. Extending this to predict two component masses was as simple as adding an extra neuron to the output layer, which suggests that it would be straightforward to extend our method to predict any number of parameters such as spins, eccentricities, etc. By incorporating examples of transient detector noise in the training set, the DNNs can also be taught to automatically ignore or classify glitches. We have only explored very simple DNNs in this first study, therefore, it is expected that more complex DNNs would improve the accuracy of interpolation between GW templates for prediction as well as the sensitivity at low SNR, while retaining real-time performance.

Based on our preliminary results, we expect \texttt{Deep Filtering} to be able to learn from and adapt to the characteristics of LIGO noise when trained with real data. The performance of this algorithm with real aLIGO data, especially in the presence of glitches and for the detection of true GW events, will be demonstrated in a following work.

Deep learning is known to be highly scalable, overcoming what is known as the curse of dimensionality~\cite{Scaling}. This intrinsic ability of DNNs to take advantage of large datasets is a unique feature to enable simultaneous GW searches over a higher dimensional parameter-space that is beyond the reach of existing algorithms. Furthermore, DNNs are excellent at generalizing or extrapolating to new data. We have shown that our DNNs, trained with only signals from non-spinning BHs on quasi-circular orbits, can detect and reconstruct the parameters of eccentric and spin-precessing compact sources that may go unnoticed with existing aLIGO detection algorithms~\cite{Huerta:2017a,Tiwari:2016,Huerta:2014,Huerta:2013a}. It is probable that our classifier is already capable of detecting even more types of signals, beyond what we have tested.

As our understanding of scientific phenomena improves and catalogs of NR simulations become available, new categories of detected and simulated GW sources can be easily added to the training datasets with minimal modifications to the architecture of DNNs. Multi-task learning~\cite{MultiTaskCNN} allows a single DNN to classify inputs into categories and sub-categories, while also performing parameter estimation for each type of signal. This means that simultaneous real-time searches for compact binary coalescence, GW bursts, supernovae, and other exotic events as well as classification of noise transients can be carried out under a single unified pipeline.

Our DNN algorithm requires minimal pre-processing. In principle, aLIGO's colored noise can be superimposed into the training set of GW templates, along with observed glitches. It has been recently found that deep CNNs are capable of automatically learning to perform band-pass filtering on raw time-series inputs~\cite{DCNN-Raw}, and that they are excellent at suppressing highly non-stationary colored noise~\cite{SpeechEnhancement} especially when incorporating real-time noise characteristics~\cite{NoiseCues}. This suggests that manually devised pre-processing and whitening steps may be eliminated and raw aLIGO data can be fed to DNNs. This would be particularly advantageous since it is known that Fourier transforms are the bottlenecks of aLIGO pipelines~\cite{2016CQGra..33u5004U}.

Powerful modern hardware, such as GPUs, ASICs, or FPGAs, are essential to efficiently train DNNs. An ideal choice would be the new NVIDIA DGX-1 supercomputers dedicated for deep learning analytics located on-site at each of the LIGO labs. However, once DNNs are trained with a given aLIGO PSD, they can be more quickly re-trained, via transfer learning, during a detection campaign for recalibration in real-time based on the latest characteristics of each detectors' noise. Deep learning methods can also be immediately applied through distributed computing via citizen science campaigns such as Einstein@Home~\cite{Einstein@Home} as several open-source deep learning libraries, including MXNet, allow scalable distributed training and evaluation of neural networks simultaneously on heterogeneous devices, including smartphones and tablets. Low-power devices such as FPGAs and GPU chips dedicated for deep learning inference~\cite{FPGA,GPUinference,InferenceEIE} may even be placed on the GW detectors to reduce data transfer issues and latency in analysis. 

DNNs automatically extract and compress information by finding patterns within the training data, creating a dimensionally reduced model~\cite{BengioScience}. Our fully trained DNNs are each only 2\textit{MB} (or 23\textit{MB} for the deeper model) in size yet encodes all the relevant information from about 2500 GW templates (about 200\textit{MB}, before the addition of noise) used to generate the training data. Once trained, analyzing a second of data takes only milliseconds with a single CPU and microseconds with a GPU. This means that real-time GW searches could be carried out by anyone with an average laptop computer or even a smartphone, while big datasets can be processed rapidly in bulk with inexpensive hardware and software optimized for inference. The speed, power efficiency, and portability of DNNs would allow rapidly analyzing the continuous stream of data from GW detectors or other astronomical facilities. 

\subsection*{Coincident Detection of GWs and EM Counterparts}

BNS inspirals have been confirmed as the engines of short gamma ray bursts (sGRBs)~\cite{BNSdet:2017,Eichler:1989,Paczynski:1986,Narayan:1992,Kochanek:1993mw,mma:2017,Piran:2013M,Lee:2010,Lee:2007N}. We expect that future detections of NSBH mergers may confirm whether these systems are also the progenitors of sGRBs, and whether rapidly rotating hypernovae are the progenitors of long duration GRBs, collapsars, etc.~\cite{sum:2009CQGra,phi:2009astro2010S}. DNNs are particularly suited for image and video processing, therefore, they can be trained to simultaneously search for GW transients and their EM counterparts using telescopes' raw image data~\cite{CNNTransients}. If the identification of an EM transient can be carried out quickly, we can interface this information with a DNN-based GW detection pipeline and vice-versa. Joint analyses of this nature will enable real-time multimessenger astrophysics searches.

Recent work suggests that space-based GW detectors such as the evolved Laser Interferometer Space Antenna (eLISA)~\cite{PAmaro:2012CQG,GairL:2013} will be able to detect stellar mass BBH systems weeks before they merge in the frequency band of ground-based GW detectors~\cite{sesa:2016PhRvL}. DNNs can be used to detect these sources in the eLISA and aLIGO frequency bands using a unified pipeline (on-board analysis may be possible in space with extremely power-efficient chips dedicated for deep learning inference). Furthermore, by training similar DNNs, low-latency classification algorithms to search for EM transients in the anticipated sky region where these events are expected to occur.

In summary, the flexibility and computational efficiency of DNNs could promote them as standard tools for multimessenger astrophysics.

\subsection*{Scope for Improvements}

One may construct a multi-dimensional template bank using available semi-analytical waveform models, and all available NR waveforms. Thereafter, one can superimpose samples of real aLIGO noise, and non-Gaussian noise transients, on these templates, and carry out an intensive training procedure with coincident time-series inputs from multiple detectors. Once this process is finished, the DNN may be used for real-time classification and parameter estimation, including sky localization, while being periodically re-trained with more gravitational waveforms and recent aLIGO noise. Time-series inputs from multiple detectors may be provided directly to the CNNs and more neurons may be added in the final layer to predict more parameters such as spins, eccentricity, time difference, location in the sky, etc. The hyperparameters of the neural networks may be tuned, and more layers may be added to further improve the performance of \texttt{Deep Filtering}.

CNNs are limited by the fact that they can only use fixed length tensors as inputs and outputs and thus require a sliding window technique in practice. On the other hand, RNNs, the deepest of all neural networks, have cyclic internal structures and are well-suited for time-series analysis since they can make decisions based on a continuous stream of inputs rather than a vector of fixed length~\cite{DL-Review}, however, they are harder to train~\cite{DifficultyRNN}. A powerful type of RNN called LSTM (Long-Short-Term-Memory)~\cite{LSTM} is capable of remembering long-term dependencies in the input sequence. Therefore RNNs~\cite{DL-Review} are ideal for processing temporal data as they can take inputs of variable lengths and have been remarkably successful at voice recognition problems~\cite{SpeechRNN}. We are developing sequence-to-sequence models with LSTM RNNs and CNNs which can be used to denoise the input time-series and produce the clean signal as output. This pre-processed data can then be fed into our \texttt{Deep Filtering} pipeline so as to further improve the sensitivity at very low SNR.

Stacking time-series datasets to produce multi-dimensional tensors can facilitate processing massive quantities of data efficiently on modern hardware, for e.g., to find signals that are very long in duration like BNS inspirals. The accuracy of the DNNs can be further enhanced by training an ensemble of different models and averaging the results for each input~\cite{DL-Book}. 

aLIGO uses a variety of independent sensors to monitor the environment and assess data quality. Many algorithms are currently used to estimate periods which must be vetoed due to disturbances that lead to a loss in detector sensitivity. Data quality information from these auxiliary channels may also be incorporated to improve robustness of signal detection and parameter estimation in the presence of glitches and for detector characterization~\cite{lauran:2015}. 

In a broader context, our results indicate that, given models or template banks of expected signals, \texttt{Deep Filtering} can be used as a generic tool for efficiently detecting and extracting highly noisy time-domain signals in any discipline.

\section{Conclusion}
\label{conc}

We have presented a novel framework for signal processing that is tailored to enable real-time multimessenger astrophysics, and which can enhance existing data analysis techniques in terms of both performance and scalability. We exposed CNNs to time-series template banks of GWs, and allowed it to develop its own strategies to extract a variety of GW signals from highly noisy data. The DNN-based prototype introduced in this article provides a strong incentive to conduct a more comprehensive investigation and optimization of DNNs to build a new data analysis pipeline based on \texttt{Deep Filtering}, trained with real detector noise, including glitches, and the largest available template banks covering the entire parameter-space of signals, to incorporate glitch classification and to accelerate and broaden the scope of GW searches with aLIGO and future GW missions. We are currently collaborating with the developers of the \texttt{PyCBC} pipeline~\cite{2016CQGra..33u5004U}, which is routinely used for GW detection both in off-line and on-line mode, to implement \texttt{Deep Filtering} as a module to increase the science reach of GW astronomy.

The known scalability of deep learning to high-dimensional data allows the use of as many GW templates as needed to train DNNs to simultaneously target a  broad class of astrophysically motivated GWs sources. More neurons may be added to encode as much astrophysical information as needed for predicting any number of parameters, and multi-task learning can unify detection and classification of different types of sources and glitches, as well as parameter estimation, with a single DNN. Therefore, we expect this approach will increase the depth and speed of existing GW algorithm allowing real-time online searches after being trained with template banks of millions or billions of waveforms.

The DNN-based pipeline can be used to provide instant alerts with accurate parameters for EM follow-up campaigns, and also to accelerate matched-filtering and detailed Bayesian parameter estimation methods. Each prediction made by the DNNs can be quickly verified by performing traditional template matching with only the templates close to the predicted parameters.  While aLIGO matched-filtering pipelines do not cover GWs from spin-precessing and eccentric BBH mergers, we have shown that DNNs were able to automatically generalize well to these signals, even without using these templates for training, having similar detection rates for all signals and small errors in estimating parameters of low mass-ratio systems. We expect that including examples of all classes of known GW signals and noise transients while training would improve the performance across the entire range of signals. We are now working on including millions of spin-precessing and eccentric templates and developing methods to train on large-scale parallel GPU clusters.

Employing DNNs for multimessenger astrophysics offers unprecedented opportunities to harness hyper-scale AI computing with emerging hardware architectures, and cutting-edge software. In addition, the use of future exascale supercomputing facilities will be critical for performing improved HPC simulations that faithfully encode the gravitational and EM signatures of more types of sources, which will be used to \textit{teach} these intelligent algorithms. We expect that our new approach will percolate in the scientific community and serve as a key step in enabling real-time multimessenger observations by providing immediate alerts for follow-up after GW events. As deep CNNs excel at image processing, applying the same technique to analyze raw telescope data may accelerate the subsequent search for transient EM counterparts. We also anticipate that our new methodology for processing signals hidden in noisy data will be useful in many other areas of engineering, science, and technology. Therefore, this work is laying the foundations to integrate diverse domains of expertise to enable and accelerate scientific discovery.

\section*{Acknowledgments}

This research is part of the Blue Waters sustained-petascale computing project, which is supported by the National Science Foundation (awards OCI-0725070 and ACI-1238993) and the state of Illinois. Blue Waters is a joint effort of the University of Illinois at Urbana-Champaign and its National Center for Supercomputing Applications. The eccentric numerical relativity simulations used in this article were generated on Blue Waters with the open source, community software, the Einstein Toolkit. We express our gratitude to Gabrielle Allen, Ed Seidel, Roland Haas, Miguel Holgado, Haris Markakis, Justin Schive, Zhizhen Zhao, other members of the \href{http://gravity.ncsa.illinois.edu}{NCSA Gravity Group}, and Prannoy Mupparaju for their comments and interactions and to the many others who provided feedback on our manuscript. We thank Vlad Kindratenko for granting us unrestricted access to numerous GPUs and HPC resources in the Innovative Systems Lab at NCSA. We are grateful to NVIDIA for their generous donation of several Tesla P100 GPUs, which we used in our analysis. We also acknowledge Wolfram Research for technical assistance and for developing the software stack used to carry out this study and draft this publication.


\bibliography{references,references2}

\end{document}